\documentclass[aps,pre,superscriptaddress,amsmath,amssymb,twocolumn]{revtex4-1}
\usepackage{graphicx}
\usepackage{amsmath}
\usepackage{hyperref}
\usepackage{color}
\usepackage{bigints}
\usepackage{mathrsfs}
\usepackage{algorithm}
\usepackage{algpseudocode}
\definecolor{darkblue}{RGB}{8,81,156}
\hypersetup{colorlinks,breaklinks,
            linkcolor=darkblue,urlcolor=darkblue,
           anchorcolor=darkblue,citecolor=darkblue}

\date{\today}

\allowdisplaybreaks

\begin{document}

\title{Thermodynamically Driven Assemblies and Liquid-liquid Phase Separations in Biology}

\author{Hanieh Falahati}
\email{hanieh.falahati@yale.edu}
\affiliation{Department of Neuroscience, Yale School of Medicine, New Haven, CT, 06510}

\author{Amir Haji-Akbari}
\email{amir.hajiakbaribalou@yale.edu}
\affiliation{Department of Chemical and Environmental Engineering, Yale University, New Haven, CT  06520}

\begin{abstract}
The sustenance of life depends on the high degree of organization that prevails through different levels of living organisms, from subcellular structures such as biomolecular complexes and organelles to tissues and organs.
The physical origin of such organization is not fully understood, and even though it is clear that cells and organisms cannot maintain their integrity without consuming energy, there is growing evidence that individual assembly processes can be thermodynamically driven and occur spontaneously due to changes in thermodynamic variables such as intermolecular interactions and concentration. Understanding the phase separation \emph{in vivo} requires a multidisciplinary approach, integrating the theory and physics of phase separation with experimental and computational techniques. This paper aims at providing a brief overview of the physics of phase separation and its biological implications, with a particular focus on the assembly of membraneless organelles. We discuss the underlying physical principles of phase separation from its thermodynamics to its kinetics. We also overview the wide range of methods utilized for experimental verification and characterization of phase separation of membraneless organelles, as well as the utility of molecular simulations rooted in thermodynamics and statistical physics in understanding the governing principles of thermodynamically driven biological self-assembly processes.
\end{abstract}

\maketitle

\section{Introduction}\label{section:intro}
According to the second law of thermodynamics,  universe is inevitably moving towards increasing its entropy, which is usually interpreted as lack of order and organization~\cite{FrenkelPhysA1999}. Living organisms, however, maintain a high level of intercellular and subcellular organization and compartmentalization by consuming energy. Subcellular organization provides cells with specialized  micro-environments for different cellular functions, while intercellular organization within multicellular organisms makes the formation and functioning of specialized tissues and organs possible. Due to this preponderance of order, it was generally thought that the emergence of order in biological cells can only occur through active processes. A new paradigm that is becoming increasingly popular recently, however, questions this widely accepted viewpoint, and argues that even though energy needs to be consumed for maintaining the integrity of biological cells and organisms, individual assembly processes within them can still occur via thermodynamically-driven phase transitions~\cite{Sear_2007}.

\begin{figure*}
  \begin{center}
    \includegraphics[width=.95\linewidth]{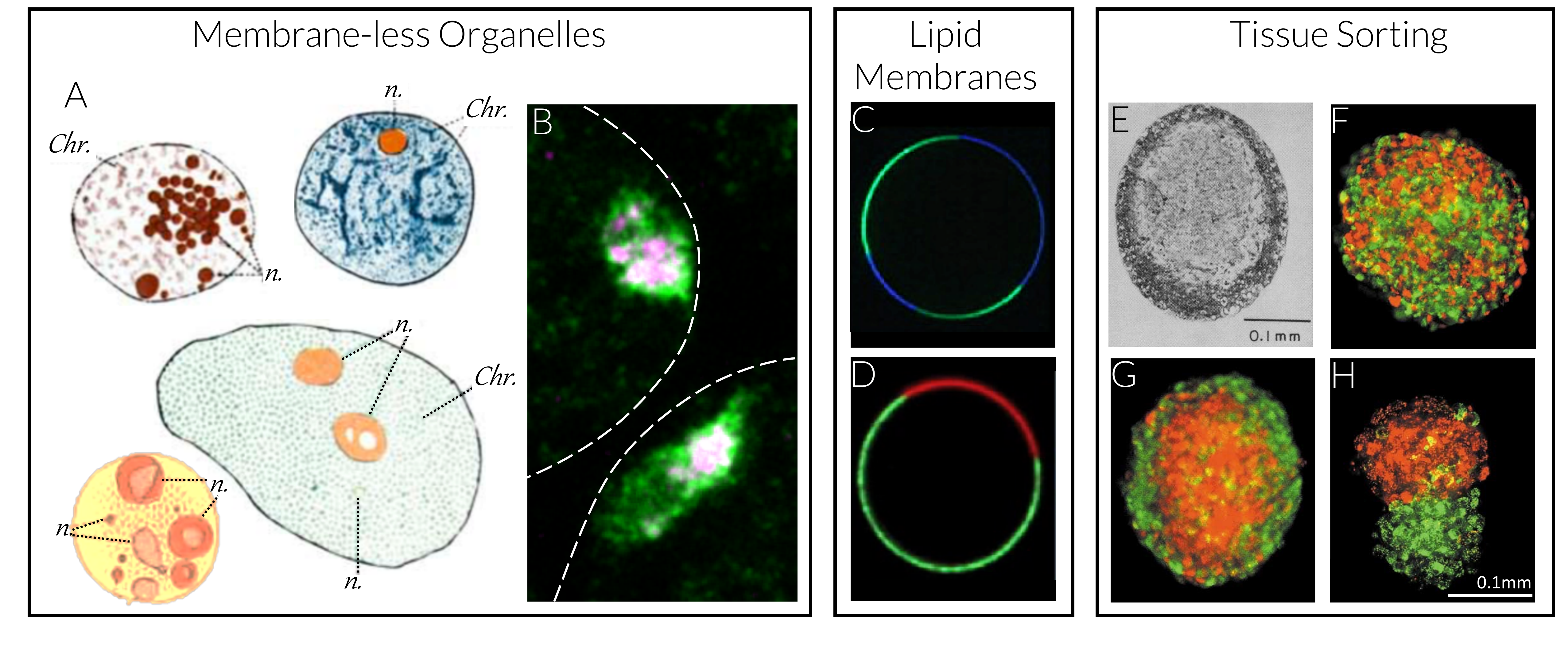}
    \caption{\textbf{Phase separation in biological systems.} The formation of membraneless organelles is proposed to be driven by LLPS. \textbf{(A)} Hand drawing of the nucleolus (\textit{n.}) in the nuclei of different cell types by Montgomery in 1898\cite{JMOR:JMOR1050150204}. Based on their shape and behavior, he noticed that the nucleoli are liquid or vicid. \textit{Chr.}: Chromatin. \textbf{(B)} Nucleolus is not homogeneous. Two fluorescently-tagged nucleolar proteins, Fibrillarin (magenta) and Modulo (green), exhibit different localization patterns within  nascent nucleoli of \emph{Drosophila} embryos. Images were obtained using expansion microscopy\cite{Chen_2015}, and the dashed lines show the boundaries of nuclei. \textbf{(C-D)} Lipid membranes are capable of separating into inhomogeneous subdomains through LLPS. Macroscopic phases of ordered and disordered liquids separated in giant unilamellar vesicle (GUV) \textbf{(C)} or vesicles derived from plasma membranes \textbf{(D)} (From \citenum{Lingwood46}). \textbf{(E-H)} LLPS is also proposed to be able to drive cellular organization. \textbf{(E)} A mixture of dissociated 5-day heart ventricle cells and 5-day liver cells of chick embryo demix into a sphere, with the liver tissue enveloping a core of heart tissue (from Steinberg 1963\cite{10.2307/1711447}). \textbf{(F-H)} Equilibrium distribution of cells expressing different levels and kinds of Cadherins. \textbf{(F)} Mixture of N5A cells expressing the same type and level of Cadherin remain intermixed and form a sphere. \textbf{(G)} Demixing of N5A cells (red) that express N-cad at levels 50\% higher  than N2 cells (green). \textbf{(H)} A mixture of B-cad (green) or R-cad (red) expressing cells separate into clusters with red cells partially capping a B-cad-expressing mass. Adopted from Duguay \emph{et al.} 2003\cite{DUGUAY2003309}.}
    \label{fig:Montgomery}
  \end{center}
\end{figure*}

For instance, consider the structural organization of biological cells, which are comprised of two types of organelles. Many intracellular organelles are membrane-bound, and their composition is maintained through active transport of molecules and ions across their surrounding membranes. Examples include endoplasmic reticulum~\cite{PorterJExpMed1945}, mitochondria~\cite{BendaArchAnatPhysiol1898} and lysosomes~\cite{CastroObregonNatEd2010}.
A second class of intracellular organelles, such as nucleoli~\cite{JMOR:JMOR1050150204, BrownPNAS1964}, Cajal bodies~\cite{CajalTrabLabInvestBiol1903}, and stress granules~\cite{GutierezBeltranPlantCell2015}, lack bounding membranes, and are instead comprised of highly concentrated assemblies of different proteins and RNAs. The question of how membraneless organelles form has fascinated biologists since the initial discovery of the nucleolus, the quintessential membraneless organelle, in the 18th century.  For instance, as early as 1898, Montgomery conducted a comprehensive investigation of nucleoli in different cell types, and presented his findings in the form of 346 hand-drawn figures (a sample shown in Figure~\ref{fig:Montgomery}A)\cite{JMOR:JMOR1050150204}. He characterized nucleoli as ''\emph{masses of varying dimensions, which may be either globular or irregular in shape, according as they are fluid or viscid in consistency}``. He further described the nucleoli to form via ''\emph{coalescence of numerous small portions of nucleolar substance}``, consistent with its fluidity\cite{JMOR:JMOR1050150204}. However, this model faded away due to advancements in cell biology and genetics, which demonstrated that  nucleoli  form around ribosomal DNA (rDNA) repeats, which are sites of active transcription and processing of ribosomal RNA (rRNA), and ribosomal biogenesis. By the end of the 20th century, membraneless organelles were commonly thought to form via active processes, commensurate with their active biological function. A number of influential works by Sear~\cite{Sear_2007}, Brangwynne~\emph{et al.}\cite{Brangwynne1729}, Li~\emph{et al.}\cite{Li:2012aa}, and Kato~\emph{et al.}\cite{Kato2012753}, however, redirected the focus to the liquid/gel nature of membraneless organelles, and the possibility that thermodynamically driven liquid-liquid phase separation (LLPS) might be responsible for their \emph{in vivo} assembly\cite{HymanRev, Banani_2017, BerryRepProgPhys2018, WeberArXiv2018, AumillerIntRevMolCellBio2014, Hyman1047, Shineaaf4382, Courchaine1603, Aguzzi2016547,WEBER201762, MITTAG20184636, Ramaswami:aa, Bah25032016}.

Another possible example of thermodynamically-driven phase separation in living systems is the reorganization of lipids and proteins within biological membranes (Figure \ref{fig:Montgomery}C-D). In principle, the biological membranes that encompass cells and bounded organelles are spatially inhomogeneous two-dimensional mixtures of a wide variety of proteins and lipids. Lipid drafts, or microdomains enriched in cholesterol and saturated lipids such as sphingomyelin, are the most widely known manifestations of such heterogeneity. They are proposed to play an important role in the function and localization of certain membrane proteins such as ion channels, and are suspected to form as a result of thermodynamically driven phase separation occurring within two-dimensional liquid-like membranes\cite{Baumgart3165,10.7554/eLife.19891,Heberle01042011,Lingwood46}. 

Phase separations in biology are not limited to intracellular processes, and have also been invoked~\cite{10.2307/1711447} to explain the organization of tissues within multicellular organisms (Figure \ref{fig:Montgomery}E-H). For instance, the dissociated embryonic cells originating from the same tissue are known to form spherical bodies commensurate with a tissue constituting a distinct mesoscopic thermodynamic phase. Furthermore, mixtures of cells originating from different tissues tend to separate into clusters rich in cells belonging to distinct tissues, and liquid-like cell aggregates with lower effective surface tensions always envelop clusters with higher effective surface tensions. This mesoscopic demixing of cells is also concentration- and composition-dependent~\cite{10.2307/1711447}. All these features are consistent with a liquid-liquid phase separation, which is thought to be mediated by the differential adhesiveness of cells originating from different tissues. This difference in adhesion propensity arises from different adhesion molecules such as Cadherins\cite{FOTY2005255,DUGUAY2003309} at the surface of cells.

It is necessary to emphasize that biological phase separations are not limited to LLPS, and can sometimes culminate in the formation of crystalline solids. Unlike liquids, molecules in crystalline solids exhibit long range order. A notable example is a process known as \emph{biomineralization}, which, for instance, results in the formation of bones and teeth~\cite{AddadiAngewChem1992}. The formation of actin filaments~\cite{HalliburtonJPhysiol1887, StraubBiochimBiophysActa1950} and microtubules~\cite{WeisenbergScience1972} is also a phase separation resulting in the formation of solids.

In addition to its suggested role in cellular organization under normal circumstances, phase separation can also be pathological~\cite{Stefani2003,Ross_2004,Forman_2004}, and diseases such as Alzheimer's\cite{HardyScience2002}, Amyotrophic Lateral Sclerosis (ALS)\cite{Shaw2001,Rowland_2001}, Parkinson's\cite{Baba:1998aa} and cataract\cite{Benedek1997CataractAA} occur as a result of the emergence of pathological assemblies within cells and tissues. Understanding the role of thermodynamics in the formation of such assemblies is key to identifying effective treatments for these medical conditions. 

{In recent years, several excellent reviews~\cite{HymanRev, Banani_2017, BerryRepProgPhys2018, WeberArXiv2018, AumillerIntRevMolCellBio2014, Hyman1047, Shineaaf4382, Courchaine1603, Aguzzi2016547,WEBER201762, MITTAG20184636, Ramaswami:aa, Bah25032016, LinBiochemistry2018} have been published on the subject of biomolecular liquid-liquid phase separation.}
This {current review has been written from a molecular thermodynamics perspective and} aims at providing a brief overview of thermodynamically driven phase separations in biological systems, with a particular focus on the role of LLPS in membraneless organelle assembly. Understanding LLPS in biological systems requires an interdisciplinary approach that is built upon the theory of phase separation rooted in classical thermodynamics and polymer physics, and applying the cutting-edge experimental and computational approaches to address the complexity of biological systems.
{This paper is aimed at providing minimal conceptual ingredients of such an exploration and is organized as follows. We dedicate Sections~\ref{section:thermo} and~\ref{section:kinetics} to discuss the thermodynamics and kinetics of phase separations in multi-component systems, respectively. Section~\ref{section:experimental} discusses experimental evidence for biological LLPS, with \emph{in vitro} and \emph{in vivo} studies highlighted in Sections~\ref{section:experimental:in-vitro} and \ref{section:experimental:in-vivo}, respectively. Section~\ref{section:computational} is dedicated to molecular simulations, and their usefulness in understanding biomolecular phase separations. Finally, we put forward some major unaddressed questions about biological self-assembly and phase separation, and discuss some potential areas of future exploration in Section~\ref{section:outlook}.}

\section{Thermodynamics and Kinetics of Phase Separation}

\subsection{Thermodynamics of Phase Separation}\label{section:thermo}
Phase separation refers to a process that occurs in multi-component mixtures, and culminates in the formation of new phases with densities and/or compositions different from the original phase. The maximum number of distinct coexisting phases that can emerge within a mixture is given by the \emph{phase rule}, which can be derived from classical thermodynamics.  According to phase rule, a mixture of $k$ nonreactive components can coexist in a maximum of $k+2$ distinct phases. In mixtures with reactive components, this maximum is decreased by $r$, the number of linearly independent chemical reactions in the system, which is equal to the rank of the stoichiometry matrix~\cite{ArisIndEngChem1963}. We will primarily  focus on the coexistence of two distinct phases, as three-- or more-- phases can only coexist over a narrower range of thermodynamic variables. In a single-component system, for instance, it will only be at the triple point where three phases can coexist with one another. 

In general, phase separation usually starts within liquid (or gaseous) mixtures. 
{One notable} exception {constitutes} multi-component crystals, such as metallic alloys~\cite{BarlowJPhysChemSolids2013} or colloidal mixtures~\cite{DijkstraPhysRevLett1998}, which can in principle separate into two or more crystalline solids with distinct compositions and symmetries.
The new phase that would emerge as a result of phase separation within a liquid mixture can, however, be a solid or a liquid. The corresponding phase transitions, which are referred to as  \emph{precipitation} and \emph{liquid-liquid phase separation}, respectively, can both occur in biological systems. Precipitation plays a pivotal role in the formation of ordered structures such as actin filaments~\cite{HalliburtonJPhysiol1887, StraubBiochimBiophysActa1950} and microtubules~\cite{WeisenbergScience1972}, as well as bio-mineralization (e.g.,~bone formation)~\cite{AddadiAngewChem1992}. It can also be pathological e.g.,~in the formation of amyloid plaques in Alzheimer's disease~\cite{HardyScience2002, LursPNAS2005}.
Precipitation is also a key separation process in structural biology, utilized for protein crystallization~\cite{McPhersonMethods2004, McPhersonActaCryst2014}. 
The focus of this paper is, however, on LLPS, which is thought to play an important role in the assembly of membraneless organelles, the formation of lipid rafts in membranes, and  cell sorting in biological tissues. {In general, the term 'liquid` usually refers to a phase that is amorphous, i.e., that its molecular structure lacks long-range translational order. The particular term utilized for describing such amorphous states of matter sometimes depends on its mechanical and transport properties, i.e.,~its relaxation dynamics. In a biological context, in particular, the term liquid usually refers only to amorphous phases that relax quickly and do not therefore withstand shear deformation. This mechanical definition is of particular interest to biology, due to the functional importance of fast dynamics within the new phase (e.g.,~for transport of ions and macromolecules). Amorphous phases that relax more slowly, however, constitute a wide range of materials, from gels to glasses, and exhibit interesting dynamical properties such as aging\cite{HodgeScience1995, UtzPRL2000, CipellettiPRL2000} (i.e.,~time dependent autocorrelation). The processes such as aggregation and physical gelation that result in the formation of such amorphous states fall into the general category of liquid-liquid phase separation. } 

\begin{figure}
  \centering
    \includegraphics[width=1\linewidth]{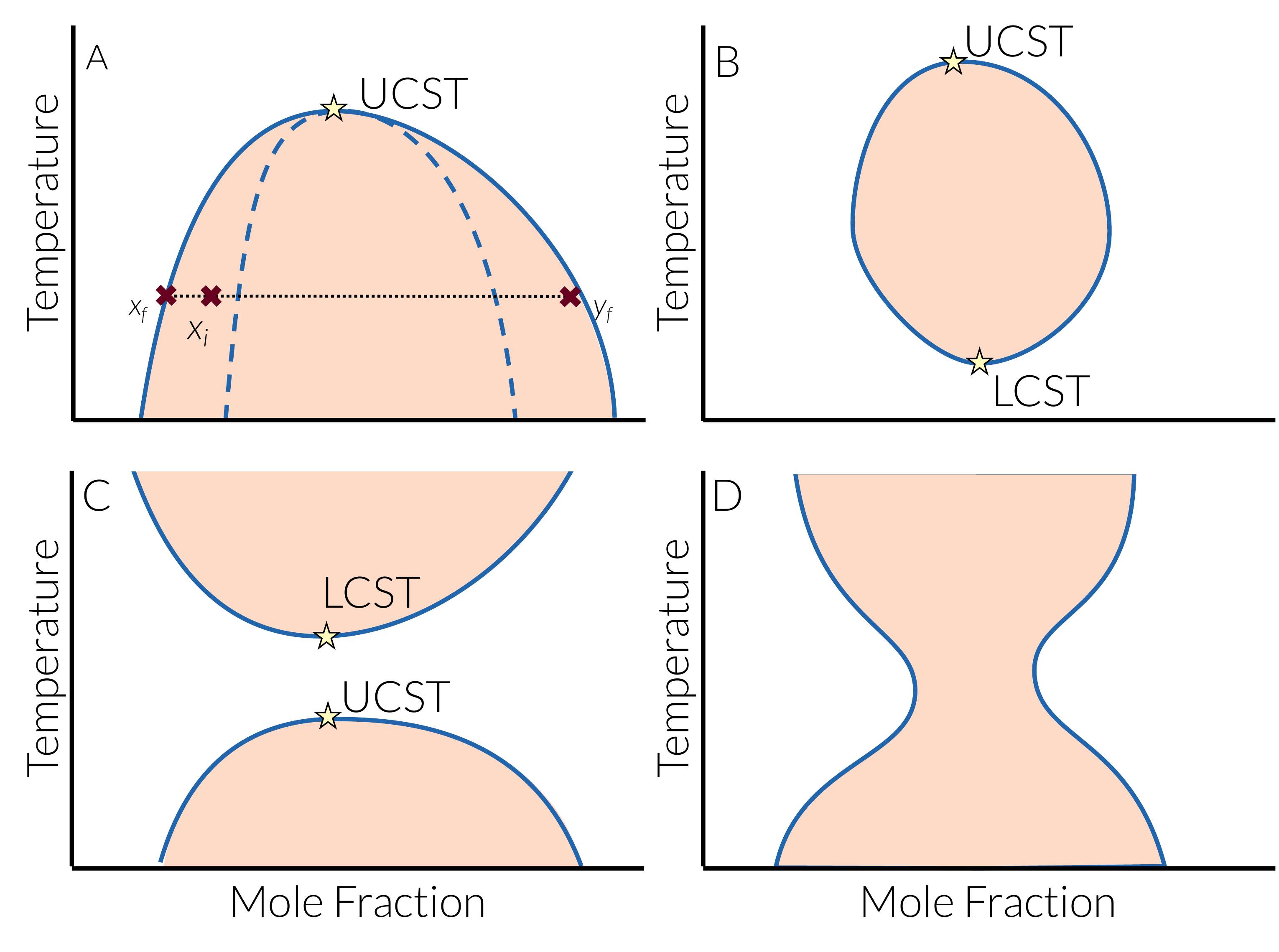}
    \caption[Types of phase diagrams]{\textbf{Schematic representation of different types of phase diagrams.} A solution that exhibits a phase separation: (\textbf{A}) with a UCST and a dome-shaped phase diagram; (\textbf{B-C}) with both an LCST and a UCST. (\textbf{B}) When the LCST is smaller than the UCST, the phase diagram will be loop-shaped, while (\textbf{C}) if LCST is larger, it will have a regular and an inverted dome; (\textbf{D}) with no critical temperature and an hourglass-shaped phase diagram. In either case, at permissive temperatures, phase separation will occur only if the mole fraction, $x^i$, is within the light orange region of the phase diagram, and will culminate in the formation of two phases with mole fractions $x^f$ and $y^f$, respectively. Note that $x^f$ and $y^f$ only depend on temperature, and not the initial composition of the mixture. Dotted line shows the tie-line and dashed line shows the spinodal line.}
    \label{fig:PhaseDiagram}
\end{figure}

In order for LLPS to be thermodynamically favored, it needs to result in a decrease in the mixture's Gibbs free energy. Let $\textbf{x}^i\equiv(x^i_1,x^i_2,\cdots,x^i_k)$ be the composition of a homogeneous liquid mixture $\mathcal{C}$, with $x^i_j$'s the mole fractions of individual components. Upon phase separation, $\mathcal{C}$ will separate into two coexisting phases, the original phase with composition $\textbf{x}^f$, and a new phase $y$ with composition $\textbf{y}^f$ (Figure \ref{fig:PhaseDiagram}A). Note that $\textbf{x}^i$ will be located on a tie line connecting $\textbf{x}^f$ and $\textbf{y}^f$, i.e.,~there will be a partition constant $0\le\lambda\le1$ so that $\textbf{x}^i=\lambda\textbf{x}^f+(1-\lambda)\textbf{y}^f$. Since the two phases $x$ and $y$ will be at equilibrium, $\pmb\mu_x(\textbf{x}^f)=\pmb\mu_y(\textbf{y}^f)$ with $\pmb\mu_\alpha\equiv(\mu_{\alpha_1},\mu_{\alpha,2},\cdots,\mu_{\alpha,k})$ the chemical potential vector for phase $\alpha=x,y$. The change in the free energy of the system as a result of phase separation will therefore be given by:
\begin{eqnarray}
\Delta{g} &=& \lambda g_x\left(\textbf{x}^f\right)+(1-\lambda)g_y\left(\textbf{y}^f\right) - g_x\left(\textbf{x}^i\right)\notag\\
&=& \lambda \textbf{x}^f\cdot\pmb\mu_x(\textbf{x}^f)+(1-\lambda)\textbf{y}^f\cdot\pmb\mu_y(\textbf{y}^f)-\textbf{x}^i\cdot\pmb\mu_x(\textbf{x}^i)\notag\\
&\overset{\pmb\mu_x(\textbf{x}^f)=\pmb\mu_y(\textbf{y}^f)}{=} & \textbf{x}^i\cdot\left[\pmb\mu_x\left(\textbf{x}^f\right)-\pmb\mu_x\left(\textbf{x}^i\right)\right]
\label{eq:delta-g}
\end{eqnarray}
Here $g_\alpha(\cdot)$ is the molar Gibbs free energy of phase $\alpha=x,y$. The described phase separation will be thermodynamically favored if $\textbf{x}^i\cdot\left[\pmb\mu_x\left(\textbf{x}^f\right)-\pmb\mu_x\left(\textbf{x}^i\right)\right]\le0$. Also, the constraints imposed on chemical potentials and the overall composition of the two phases imply that $\textbf{x}^f$, $\textbf{y}^f$ and $\lambda$ can be uniquely determined from the temperature, pressure and composition of the original mixture. In general, it is possible to obtain empirical or semi-empirical expressions for $\pmb\mu_\alpha$, which can then be utilized to predict the possibility-- or lack thereof-- phase separation for any given mixture. For the special case of a binary system with components $A$ and $B$, the compositions of coexisting phases can be uniquely determined for a given temperature and pressure, and the phase separation will be thermodynamically favored if:
\begin{eqnarray}
\int_{x_A^i}^{x_A^f}\left[\frac{x_A^i}{1-x_A^i}-\frac{x_A}{1-x_A}\right]\left(\frac{\partial\mu_A}{\partial x_A}\right)_{P,T}dx_A < 0 \label{eq:delta-g-binary}
\end{eqnarray}
{Eq.~(\ref{eq:delta-g-binary}) follows from (\ref{eq:delta-g}) and the Gibbs-Duhem equation, i.e.,~$d\mu_B=-x_Ad\mu_A/(1-x_A)$.}
Eqs.~(\ref{eq:delta-g}) and (\ref{eq:delta-g-binary}) describe the necessary thermodynamic conditions for phase separation, without telling us anything about its underlying physics. The latter can be attained by decomposing $\Delta{g}$ into its enthalpic and entropic contributions, which correspond to the strength of intermolecular interactions, and the number of accessible microstates, respectively.  Letting $\Delta{g}=\Delta{h}-T\Delta{s}$ and assuming that neither $\Delta{h}$ nor $\Delta{s}$ are strong functions of temperature, one can imagine the following  scenarios. In the case of $\Delta{h}>0$ and $\Delta{s}<0$, phase separation cannot occur under any circumstances, and the system will remain mixed at all conditions. If $\Delta{h}<0$ and $\Delta{s}>0$, the system will always be demixed and a homogeneous mixture will be unstable. We will consider the other two scenarios in which the system can remain homogeneous, or phase separate depending on thermodynamic conditions. 

The most common scenario is when $\Delta{h}<0$ and $\Delta{s}<0$, and corresponds to situations in which molecules within the demixed phases experience stronger intermolecular interactions. The system, however, experiences a decline in the number of accessible microstates due to demixing.  This usually occurs when there are unfavorable intermolecular interactions between some components within the original mixture, and results in the separation of the components that do not ''like`` each other into distinct phases. Such separation, however, eliminates all the microstates in which molecules of different types are randomly distributed within a homogeneous liquid, and results in a decrease in entropy. The free energetic penalty associated with such entropic loss, $-T\Delta{s}$, will be proportional to temperature. Therefore, beyond a critical temperature known as \emph{upper critical solution temperature (UCST)}, demixing will become thermodynamically unfavorable, and the system can only exist in a mixed state. 

Figure ~\ref{fig:PhaseDiagram}A depicts the prototypical phase diagram for a binary mixture with a UCST. According to the phase rule, temperature and the composition of the original homogeneous mixture are sufficient for determining whether phase separation occurs, and if so for predicting the compositions of coexisting liquids.  As can be seen in Figure ~\ref{fig:PhaseDiagram}A, at concentrations outside the shaded dome, the system will remain fully mixed. What is notable about binary-- or pseudo-binary-- mixtures is that changing the concentration within the dome will not affect the compositions of the coexisting liquids, and will only alter the partition constant, $\lambda$. In biological systems, examples of UCST have been observed \emph{in vitro} for the low-complexity domain of the RNA binding protein FUS \cite{Burke2015231}, a disordered Nauge protein \cite{Nott2015936}, and lipid bilayers \cite{PhysRevLett.91.245701}, and \emph{in vivo} for the nucleolar proteins Fibrillarin, Nopp140, RNA polymerase I, and Pitchoune\cite{Falahati07022017}.

The other less common situation occurs when the entropy of demixing is positive, i.e.,~when the mixture is more ordered than the separated phases. This can occur for one of the following reasons.  One possibility is the existence of very strong and/or highly directional intermolecular interactions between non-identical components in the original mixture. In a binary mixture, for instance, this can occur if hydrogen bonds can only form between $A$ and $B$ molecules. At low temperatures, demixing will be thermodynamically unfavored since the enthalpic cost of breaking such strong interactions between unlike molecules cannot be compensated by the entropic gain due to demixing. Beyond a critical temperature known as the \emph{lower critical solution temperature (LCST)}, however, the entropic term will become dominant and the system will phase separate. {The existence of LCST can therefore be attributed to effective interactions that depend on temperature~\cite{LinBiochemistry2018}.} Such mixing-induced ordering will eventually fade at sufficiently high temperatures due to the weakening of the original directional interactions, so a UCST will also exist, as depicted in Figure~\ref{fig:PhaseDiagram}B. In other words, when the LCST temperature is smaller than UCST, the dome of Figure~\ref{fig:PhaseDiagram}A will be replaced with a loop. This behavior has been observed \emph{in vitro} for a spindle-associated protein, BuGZ \cite{Jiang2015108}.

Another type of mixing-induced ordering can occur for polymeric solutions close to the vapor-liquid critical point of the pure solvent. Under such circumstances, critical fluctuations in the solvent will decrease the number of configurations accessible to polymeric chains, henceforth culminating in a negative entropy of mixing. This scenario, which is uncommon in biological systems, results in phase diagrams shown in Figure~\ref{fig:PhaseDiagram}C and ~\ref{fig:PhaseDiagram}D~\cite{prausnitz1998molecular} depending on the separation between the low-temperature demixing region and the critical temperature of the solvent.

Despite this lack of universality in {how} thermodynamically driven phase separations are affected by changes in temperature, they are  distinct from active assembly processes in that they are all \emph{reversible}. In other words, active assembly processes are driven by irreversible enzymatic reactions and proceed at higher rates when temperatures are higher. {This means that} changing temperature will only impact their {kinetics}, and not their {overall} direction. Thermodynamically driven assemblies, however, can be reversed by changing temperature, e.g.,~increasing it in the case of a UCST or decreasing it in the case of an LCST. Such reversibility has been shown to be pivotal in determining whether a  particular \emph{in vivo} assembly process  is thermodynamically driven or active, e.g.,~by monitoring its response to oscillations in temperature\cite{Falahati07022017}.

{At the end of this section, it is necessary to emphasize that while we have only discussed phase separation in mixtures, both liquid-liquid\cite{GlosliPRL1999, KatayamaNature2000, PalmerNature2014, PalmerJChemPhys2018} and solid-solid\cite{MinomuraJPhysChemSol1962, BolhuisPhysRevLett1994, JacobsScience2001, WinkelJChemPhys2008, HajiAkbaricondmat2011, HajiAkbariDQC2011, PengNatMater2015} transitions can also occur in single-component molecular\cite{MinomuraJPhysChemSol1962, GlosliPRL1999, KatayamaNature2000, JacobsScience2001, WinkelJChemPhys2008, PalmerNature2014, PalmerJChemPhys2018} and colloidal\cite{BolhuisPhysRevLett1994, HajiAkbaricondmat2011, HajiAkbariDQC2011, PengNatMater2015} systems.  Unlike phase separation in mixtures, the coexisting phases in single-component systems can only differ in density or symmetry (i.e.,~the arrangement of their constituent molecules). The ability of a pure substance to exist in distinct crystalline and amorphous forms is referred to as \emph{polymorphism} and \emph{polyamorphism}, respectively, and has been extensively studied  in experiments\cite{MinomuraJPhysChemSol1962, KatayamaNature2000,  JacobsScience2001, WinkelJChemPhys2008, PengNatMater2015} and simulations\cite{BolhuisPhysRevLett1994, GlosliPRL1999,  HajiAkbaricondmat2011, HajiAkbariDQC2011, PalmerNature2014, PalmerJChemPhys2018}.}

%
%

\begin{figure}
  \begin{center}
    \includegraphics[width=1\linewidth]{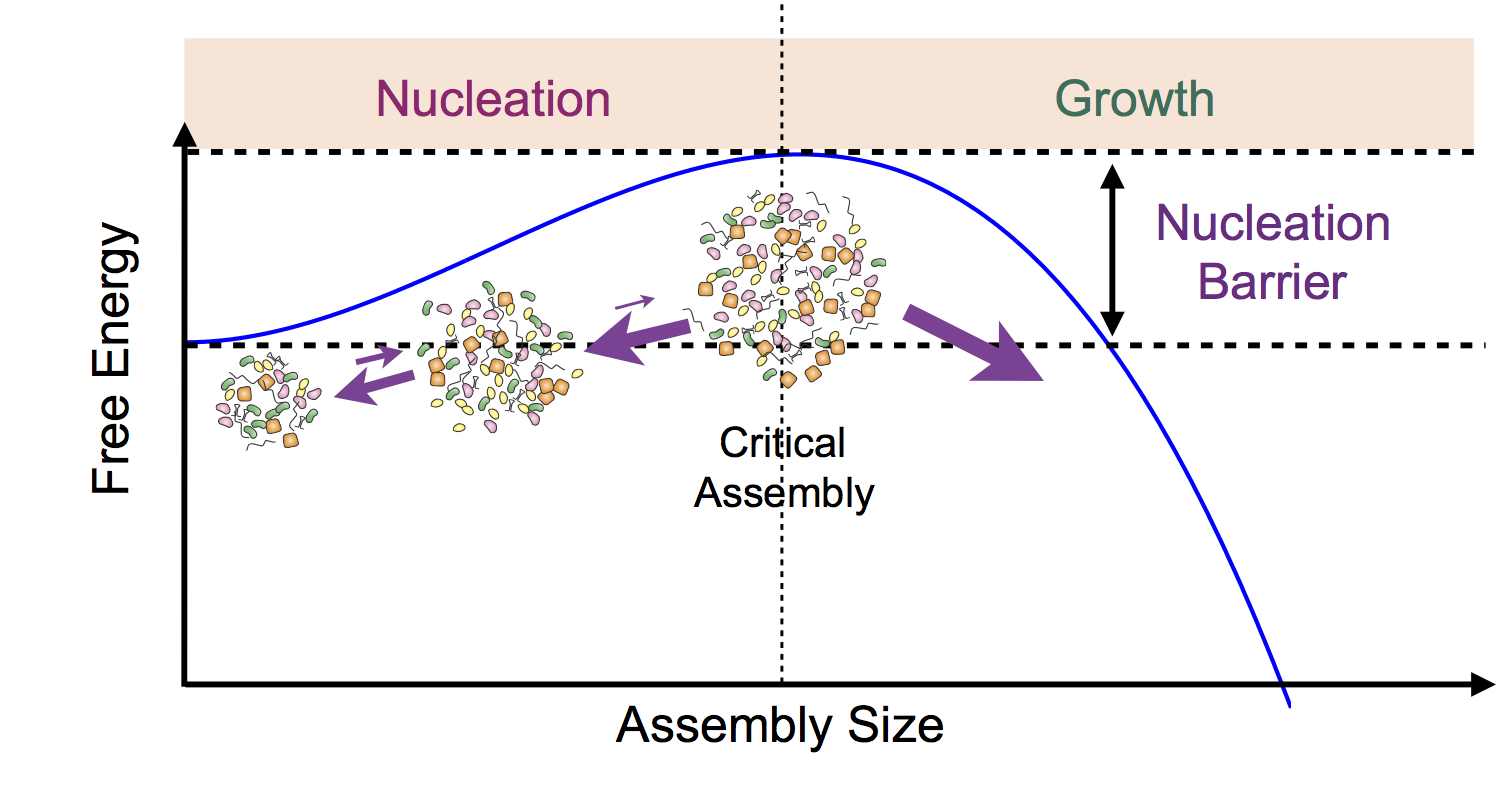}
    \caption[Nucleation and growth in phase separation processes] {\textbf{Nucleation and growth in phase separation processes.} A first-order phase separation starts with nucleation during which tiny assemblies of the new phase emerge within the old phase. Smaller assemblies are thermodynamically unstable due to their large specific surface areas, and are therefore more likely to shrink unless they reach a critical size.  If the nucleation barrier is large, then the formation of new assembly becomes nucleation-limited. Otherwise, this process becomes growth limited.}
    \label{fig:Nucleation}
  \end{center}
\end{figure}

As a first-order phase transition, LLPS can occur via two distinct mechanisms, depending on the magnitude of the thermodynamic driving force $\Delta{g}$. For small $\Delta{g}$'s, e.g.,~close to the phase boundaries of Figure~\ref{fig:PhaseDiagram}, demixing will occur through \emph{nucleation and growth} (Figure~\ref{fig:Nucleation}). During nucleation, which is an activated stochastic process, instantaneous thermal and compositional fluctuations result in the formation of small nuclei of the new phase within the old mixture. Such nuclei will be thermodynamically unstable and will therefore be more likely to melt due to their large specific surface areas, unless they are larger than a critical size. The free energetic cost of forming such a critical nucleus,  $\Delta{G}_{\text{nuc}}$, is referred to as the \emph{nucleation barrier}. The likelihood that a critical nucleus will form within a mixture is proportional to $e^{-\Delta{G}_{\text{nuc}}/kT}${, with $k$ the Boltzmann constant. Therefore a larger $\Delta{G}_{\text{nuc}}$ will imply a lower likelihood for the formation of a critical nucleus.} Upon its formation, however, a critical nucleus can grow until the system reaches the predicted partition constant $\lambda$, and the growth timescale will scale with transport properties of the original mixture. The separation between the nucleation and growth timescales will be larger when $\Delta{G}_{\text{nuc}}\gg kT$, and phase separation will become a nucleation-limited rare event. The closer the nucleation barrier is to $kT$, the smaller will such a separation of timescales be. In the limit of $\Delta{G}_{\text{nuc}}\rightarrow kT$, LLPS will be growth-limited. Whenever the nucleation rate is large, e.g.,~in the growth-limited regime, multiple nucleation events will occur within a single mixture, and the completion of LLPS will proceed through {several simultaneous non-equilibrium} process{es\cite{AvramiJChenPhys1939}. One such process that occurs as a result of diffusion within the continuous phase is} known as \emph{coarsening} or \emph{Ostwald ripening}~\cite{VoorheesJStatPhys1985} in which larger nuclei will grow in the expense of smaller nuclei that melt due to their lower thermodynamic stability. {Another important process is called \emph{coalescence} in which smaller nuclei collide and join to form a larger nucleus.}

This simplified picture is the essence of \emph{classical nucleation theory (CNT)}~\cite{VolmerZeitsfPhysik1934, BeckerAnnalenderPhysziks1935}, which is the most widely used quantitative framework for understanding nucleation. CNT assumes that nucleation is a single-step process, and a steady-state distribution of precritical nuclei is established at timescales that are orders of magnitude shorter than the nucleation time. According to CNT, nucleation barrier will be given by:
\begin{eqnarray}
\Delta{G}_{\text{nuc}} &=& \frac{16\pi\sigma^3}{3|\Delta{g}|^2}
\end{eqnarray}
with $\sigma$ the interfacial tension between the two phases, which only depends on temperature and the compositions of the two coexisting phases. It can be generally stated that  the nucleation barrier is expected to decrease upon increasing $\Delta{g}$.  It is necessary to emphasize that the key assumptions of CNT might be violated in some systems, and alternatives of CNT, such as multi-step nucleation\cite{GebauerChemSocRev2014, ZahnChemPhysChem2015}, have been formulated in the literature.  

Like other first-order phase transitions, nucleation in LLPS can be homogeneous or heterogeneous. In homogeneous nucleation, the critical nucleus forms endogenously within the mixture, while in heterogeneous nucleation, an external entity such as an insoluble impurity provides a template for nucleation, and results in a decrease in the size of the critical nucleus and the magnitude of the nucleation barrier. Heterogeneous nucleation generally occurs at higher rates, and is {a means of} exerting spatiotemporal control on phase separation in biological systems~\cite{Falahati2016277}. An extension of CNT for heterogeneous nucleation was proposed by Turnbull~\cite{TurnballJCP1950}, and predicts that $\Delta{G}_{\text{het}} = f(\theta_c)\Delta{G}_{\text{homo}}$. Here $f(\theta_c)$ is called the \emph{potency factor} and depends on $\theta_c$ the contact angle between the two phases and the external surface. {From a physical perspective, potency factor is a measure of differential attractiveness of the interaction between the nucleating surface and the two liquids\cite{HajiAkbariJCP2017}. Lower potency factors correspond to situations in which the external surface has a higher propensity for the new phase, which results in smaller contact angles, and considerably larger nucleation rates at identical $\Delta{g}$'s.}

Nucleation and growth is the mechanism of demixing when the original mixture is metastable, i.e.,~when an arbitrary concentration fluctuation will always be dampened unless its amplitude is sufficiently large (e.g.,~in a critical nucleus). Another regime, however, is possible in which the mixture becomes mechanically unstable. From a thermodynamic perspective, this will occur if the Hessian of $g(\textbf{x})$-- {or its generalized second-order derivative with respect to its independent arguments}-- has at least one negative eigenvalue. {For a binary system, this will imply a situation in which $g''(x_A)<0$.}
 Under such circumstances, concentration  fluctuations with sufficiently large wavelengths  will be amplified, which will in turn drive the system into a demixed state. Several mean-field theories have been developed to describe this phenomena, which is typically known as \emph{spinodal decomposition}~\cite{CahnJChemPhys1958, CahnJChemPhys1959, CahnJChemPhys1959p}. Within these theories, it is predicted that a transition from the nucleation-and-growth regime to the spinodal regime occurs at a spinodal line (Figure \ref{fig:PhaseDiagram}A), which is the loci of inflection points of $g(\textbf{x})$. In the case of binary mixtures, this will correspond to loci of $g''(x)=0$. {During spinodal decomposition, fluctuations with different wavelengths get exponentially amplified, but with different time constants.  As a result, the morphology of the system is dominated by the fastest growing wavelength, resulting in a characteristic length scale for phase separated domains.} There is extensive experimental~\cite{HuangPRL1974} and computational~\cite{MruzikJChemPhys1978, AbrahamPRL1982, KochPhysRevA1983} evidence for spinodal decomposition. {Spinodal decomposition has been experimentally observed in the phase separation of Ddx4 protein\cite{LinJMolLiq2017}.} Due to the mean-field approximation inherent in the spinodal model, however, it is not trivial to pinpoint the boundary at which the phase separation mechanism transforms from nucleation and growth to spinodal decomposition.

\section{Phase Separations in Biological Systems: Experimental Investigations}
\label{section:experimental}
As mentioned in Section \ref{section:intro}, phase separation can occur in many different biological settings. The resurgent interest in the role of LLPS in biological assembly can, however, be primarily attributed to their potential role in the formation of membraneless organelles. Understanding the mechanisms by which membraneless organelles form, is particularly important as these cellular bodies carry out several essential cellular functions~\cite{Holehouse2018Biochem}. In the nucleus, for instance,  Cajal bodies form at snRNA loci and constitute the sites at which the splicing machinery is assembled~\cite{WRNA:WRNA1139, Nizami01072010}. Histone-locus bodies contain the macromolecular machinery necessary for the transcription and processing of histone mRNAs\cite{Nizami01072010}, and nucleoli are the sites of ribosomal biogenesis and several other important cellular functions. In the cytoplasm, stress granules and processing bodies (P-bodies) form in response to cellular stresses, and limit the translation of mRNAs~\cite{Li361}. Each of these organelles has a distinct composition; they are rich in RNAs and proteins necessary for their unique function, and lack the components that might  inhibit such roles\cite{Li:2012aa,Heyn2016,Langdoneaar7432}. Furthermore, these organelles are not structurally homogeneous, and the macromolecules within them tend to organize into smaller domains. For instance, depending on the organism, nucleolus can be comprised of two to three micro-domains that specialize in different steps of ribosomal biogenesis (Figure \ref{fig:Montgomery}A-B){\cite{JMOR:JMOR1050150204,RecherJUltraMolStructR1969, FERIC20161686, LinNewJPhys2017}}. Understanding the underlying mechanisms for the formation of such intricate, specialized structures, and the regulation of the timing and location of their assembly, is therefore critical to our ability to understand the biological functions of such organelles. 

\begin{figure}[h]
\centering
	\includegraphics[width=.3\textwidth]{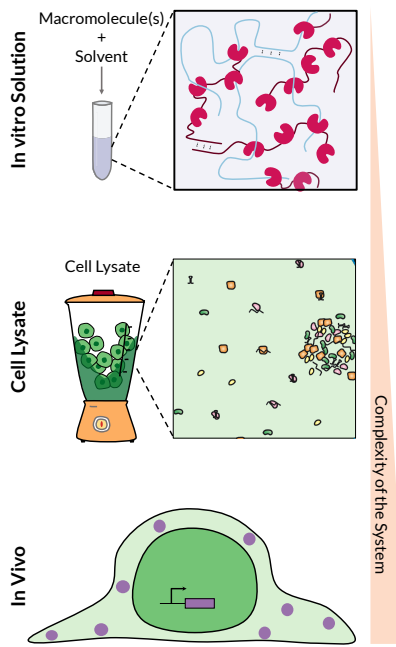}
	\caption{\label{fig:InBiology}
	{\textbf{Experimental investigations of liquid-liquid phase separation in biomolecular systems.} \emph{In Vitro} systems are typically comprised of a small number of macromolecules, such as RNAs and/or proteins, dissolved in an aqueous solution, and, due to their well-defined compositions, are ideal for examining the effect of different thermodynamic factors on the phase behavior. Furthermore, \emph{in vitro} studies have made considerable contributions to our current understanding of the molecular properties that drive phase separation, as well as the effect of pathological mutations on the demixing propensity. In order to capture the inherent complexity of biological systems, a number of \emph{in vitro} investigations have used cell lysates to study LLPS. Finally, several studies have directly focused on LLPS and membraneless organelle assembly \emph{in vivo}, in order to account for full complexity of biological systems. Such studies examine three main aspects of membraneless organelles: (i) mechanical and transport properties, (ii) the effect of thermodynamic variables, such as concentration and temperature, on LLPS, and (iii) kinetics and spatiotemporal regulation of membraneless organelle assembly. See text for more details.
	}}
\end{figure}

In addition to their significant role in normal cellular functions, membraneless organelles can exhibit anomalous morphologies in several pathological disorders. For instance, a hallmark of cancer cells is an increase in the number and size of their nucleoli, a fact that was described as early as 1896 by Giuseppe Pianese\cite{pianese1896beitrag,Ruggerope38}. 
 Moreover, mutations in stress granule proteins such as FUS have been associated with neurodegenerative diseases such as ALS or frontotemporal dementia (FTD) in which pathological aggregates form\cite{Li361, Ramaswami:aa, Patel20151066, Aguzzi2016547}. This also underlines the necessity for a better mechanistic understanding of the driving forces and the regulatory mechanisms of the membraneless organelle assembly.  
 
 In this section we describe the experimental approaches commonly used for studying phase separation in biological systems (summarized in Figure \ref{fig:InBiology}). In each case we describe what is measured by each technique, how the results support the LLPS model, and also the limitations of each approach.

\subsection{\emph{In Vitro} Investigation of LLPS}\label{section:experimental:in-vitro}
The propensity of protein/salt water mixtures to phase separate \emph{in vitro} is not new in biology. For instance, the phase diagrams of globular proteins such as lysozyme, and $\gamma_{II}$-crystallin dissolved in aqueous NaCl solutions have been studied since 1970's, and their demixing  into liquid-like, amorphous and crystalline phases at different conditions has been reported~\cite{Ishimoto:1977aa, Thomson01101987}. Such studies of phase separation were primarily conducted with the aim of understanding and optimizing protein crystallization, which is an important technique in structural biology, and a prerequisite for determining the three-dimensional structures of proteins using X-ray crystallography. In this context, liquid-like and amorphous assemblies of proteins are unwanted products of processes  aimed at producing protein crystals. 

With resurgent interest in the phase separation model and the possible role of LLPS in membraneless organelle assembly, similar \emph{in vitro} assays are now utilized for studying the condensation of proteins and/or RNAs that localize to such \emph{in vivo} assemblies.  These studies primarily focus on mapping out the phase diagrams of mixtures with one-- or a few-- types of macromolecules, by varying temperature, pH and ionic strength of the solvent, and the concentrations of the constituent macromolecules. Such \emph{in vitro} approaches have proven invaluable in assessing the sensitivity of the thermodynamics and kinetics of phase separation to changes in molecular properties, and comparing the phase separation propensity of different proteins under well-defined conditions. Such molecular properties, for instance, can be tuned by mutating different domains and residues of the phase-separating proteins, or by engineering biology-inspired modules e.g.,~by introducing sequence repeats~\cite{Li:2012aa,Quiroz:2015aa,Burke2015231, Banani_2016, Pak_2016,Zhang_2015,doi:10.1021/acs.langmuir.6b02499, Lin17112017,Li_2018}. These approaches have been widely used in understanding the role of mutations in proteins that can form pathological aggregates. Mutations in the coding regions of certain proteins such as FUS \cite{Murakami2015678,Patel20151066}, TDP-43\cite{Conicella_2016} and hnRNPA1\cite{Molliex2015123}, which are all observed in pathological disorders, are shown to change the boundaries of the phase diagram, as well as the kinetics of phase separation (referred to as aging). {Other \emph{in vitro} studies examine the effect of introducing reactions that result in post-translational modifications of proteins, such as phosphorylation, methylation or ubiquitination, on the phase  boundaries\cite{Bah25032016,Aumiller-Jr:2016aa,Monahane201696394,DAO2018965,Milovanovic604}. The effect of post-translational modifications has been observed in the assembly of proteins such as Coilin~\cite{Toyota2010}, nucleolar proteins (reviewed in Ref.~\citenum{doi:10.4161/nucl.2.3.16246}), MEG proteins\cite{10.7554/eLife.04591}, eIF-2$\alpha$\cite{Kedersha1431},  FUS\cite{Monahane201696394}, Ddx4\cite{Nott2015936}, UBQLN2\cite{DAO2018965}, and LAT\cite{Su595}, and was also used to engineer  synthetic RNA-protein assemblies \emph{in vitro}\cite{Aumiller-Jr:2016aa, Lin17112017}.}

    \emph{In vitro} studies have been particularly pivotal in identifying the types of biological macromolecules that can  phase separate under physiological conditions~\cite{MITTAG20184636}. {Phase separation has been observed in solutions of one or more proteins, mixtures of RNA and protein\cite{Molliex2015123}, solutions of RNA\cite{Jain_2017}, and mixtures of protein and liposomes\cite{Milovanovic604}. These phase separating biomolecules can be classified into two distinct categories.} The first class is comprised of those with multivalent interactions. Similar to patchy colloids {(i.e.,~colloids with multiple interaction sites at their surface)}, they typically contain repetitive modules that can attract and form non-covalent bonds with other modules in their binding partners. For proteins, these repetitive modules are often folded domains. Examples include the signaling protein WASP, which has multiple proline-rich motifs that can bind to the three SH3 domains of Nck \cite{Li:2012aa}, or proteins with multiple RNA binding domains. {In the latter case, RNA can function as a cross-linker between the proteins that contain multiple RNA binding domains.} The second class is comprised of proteins with intrinsically disordered regions (IDRs), which have also been observed to phase separate \emph{in vitro}~\cite{Kato2012753}. Such regions lack well-defined three-dimensional structures and are capable of forming multiple weak interactions with other molecules. Polypeptides that are exclusively comprised of IDRs are typically referred to as \emph{intrinsically disordered proteins (IDPs)}~\cite{TOMPA2012509}. {It is necessary to emphasize that these two categories are not mutually exclusive, as many phase separating proteins contain both repetitive modules and IDRs.} In this paper, we will mostly focus on IDPs and IDR-containing proteins and the computational approaches to study their ability to undergo phase separation.

Although such \emph{in vitro} studies have been pivotal in laying out of basic understanding of biological condensates, they are incapable of properly accounting for the complexity of \emph{in vivo} systems that are comprised of millions of different components. This is due to the fact that adding even one extra component to a mixture can potentially result in a change in the phase boundary in favor of mixing or demixing of original components, and can even cause the emergence of new liquid or solid phases that would otherwise not form in the original mixture. For instance, adding nucleophosmin to a mixture of fibrillarin in buffer will result in the emergence of a new liquid phase~\cite{FERIC20161686}. Therefore, it is not clear whether the macromolecules that phase separate \emph{in vitro} would also condense inside cells in the presence of millions of other components.

With the aim of better representing \emph{in vivo} systems, several groups have conducted \emph{in vitro} assays of successively complex multi-component mixtures. For instance, in a seminal work in Rosen's lab, the twelve components of  the T-cell receptor signaling pathway were reconstituted \emph{in vitro}, which then segregated into a multi-component assembly containing the kinase and depleted of the phosphatase, and a second aqueous phase of the phosphatase~\cite{Su595}.  Similarly, the phase separation of the six components of the post-synaptic density was studied \emph{in vitro}\cite{ZENG20181172}. Interestingly, both these reconstitutions involve lipid bilayers in order to mimic the \emph{in vivo} localization of the constituent proteins, which occurs through establishing attractive interactions between the transmembrane and cytoplasmic proteins. In both cases, only the thermodynamics of phase separation was examined. It is, however, entirely plausible that the surface of the lipid bilayer can act as a heterogeneous nucleation site for the formation of the respective assemblies, and therefore play a role in the spatiotemporal regulation of the formation of such condensates.

The \emph{in vitro} systems that are the most similar to intracellular conditions are cellular extracts, which contain the bulk of components present \emph{in vivo}. For instance, Hankock showed that expanding intact nuclei in a hypotonic medium will result in the disassembly of the nucleoli, and the effect can be reversed by adding crowding agents\cite{HANCOCK2004281}. This is consistent with the reversibility of a thermodynamically-driven phase separation. Later, Kato \emph{et al.} showed that exposing cell or tissue lysates to a chemical known as biotinylated isoxazole will induce the formation of hydrogels that include many RNA-binding proteins. These proteins mostly contain low complexity domains, or IDRs, which are proposed to drive phase separations \emph{in vivo}~\cite{Kato2012753}. Similarly, exposing \emph{Drosophila} egg chambers to  a saline solution results in the emergence of novel dynamic nuclear bodies that share several features of other membraneless organelles, such as rapid exchange of components~\cite{doi:10.4161/nucl.2.5.17250}. While these experiments demonstrate the ability of multi-protein mixtures or extracts to undergo phase separation, they also indicate the strong sensitivity of the phase boundary to small changes in composition.
Therefore, while \emph{in vitro} studies provide invaluable information about the underlying molecule driving forces for biopolymeric phase separation, \emph{in vivo} studies are necessary to determine whether the assembly of individual proteins and RNAs in living cells is truly an LLPS process.

\subsection{\emph{In Vivo} Investigation of LLPS}\label{section:experimental:in-vivo}
The complex nature of living cells, which are comprised of a large number of reactive components, makes testing the validity of the LLPS model  \emph{in vivo} extremely challenging. In recent years, several studies have attempted to address this grand challenge by examining: (i) the mechanical and transport properties of intracellular assemblies, (ii) their sensitivity to changes in thermodynamic variables, and (iii) the \emph{in vivo} kinetics of the self-assembly process.

The majority of earlier \emph{in vivo} studies of LLPS focus on examining the mechanical and transport properties of membraneless organelles~\cite{BERGERONSANDOVAL20184754, MITREA20184773}. For instance, Brangwynne \emph{et al.} reported the liquid-like behavior of P-granules, such as their propensity to wet external surfaces, and their ability to fuse with one another~\cite{Brangwynne1729}. Subsequent studies on other intracellular assemblies, including nuclear bodies\cite{JMOR:JMOR1050150204,Brangwynne15032011,Kim347955,Nizami2012}, and other cytoplasmic RNA granules\cite{Zhang_2015,Franzmanneaao5654,Gopal201614462,Jain_2017} demonstrated  that this ability to fuse is shared by all these organelles. In addition, the dynamic nature of such organelles has been extensively studied by measuring the mobility of components that localize to those assemblies. This can, for instance, be achieved by tagging the macromolecule of interest with a fluorescent probe, and  measuring the recovery rate after its photobleaching. It has been observed that the tagged macromolecules that localize into membraneless assemblies can rapidly move in and out, or within such assemblies~\cite{Phair:2000aa, Strom_2017, Falahati07022017}. However, not all components of the membraneless organelles exhibit this rapid dynamic behavior. For instance, the core of stress granules, and the MEG protein that constitute the shell of P-granules are stable structures with amorphous solid- or crystal-like dynamics~\cite{Jain:aa,Putnam245878}.

In addition to these mechanical properties, the sensitivity of such assemblies to changes in thermodynamic variables have also been examined \emph{in vivo}. As discussed in section \ref{section:thermo}, the thermodynamic driving force for phase separation is determined both by the composition of the mixture, and temperature. Most \emph{in vivo} studies examine how phase separation is impacted by changes in concentration, and demonstrate that changing the concentration of the phase-separating component, either genetically~\cite{WippichCell2013, Pak_2016, BerryPNAS2015} or optically~\cite{ShinCell2017}, will modulate the phase separation process. Also, globally increasing the concentration of macromolecules in HeLa cells by exposing them to a hypotonic solution, or decreasing the temperature results in the self-assembly of  a germ granule protein, Ddx4\cite{Nott2015936}. Interestingly, locally increasing the concentration of specific nucleolar proteins, by tagging them individually with LacI and targeting them to LacO repeats, can enrich the nucleolar proteins not tagged with LacI~\cite{Kaiser1713}. Together, these studies confirm the concentration dependence of the assembly, as expected for a thermodynamically-driven phase separation. Yet, active processes also exhibit a similar dependence of rate on concentration, rendering the findings of such studies inconclusive.

Recently, an \emph{in vivo} assay was developed by Falahati and Wieschaus that allows for distinguishing between the LLPS model and an active assembly process for bona fide organelles~\cite{Falahati07022017}. This assay is based on the two thermodynamic properties that collectively distinguish LLPS from an active assembly model, namely the temperature dependance and reversibility. This assay was utilized to study the mechanism by which six nucleolar proteins localize to the nucleolus. Using this approach they demonstrated that  Fibrillarin, Nopp140, Pitchoune and RNA polymerase I, condense at low temperatures and dissolve at ambient temperatures in \emph{Drosophila} embryos, confirming a reversible phase separation with a UCST that is responsive to rapid changes in temperature. Interestingly, not all nucleolar proteins in this study followed this behavior; During the initial growth stage, the assembly of Nucleostemin1 and Modulo, the fly homologue of Nucleolin, was inhibited at low temperatures, did not show reversibility in response to oscillations in temperature, and became insensitive to changes in temperature when a maximum size (saturation level) was achieved. This is inconsistent with the LLPS mechanism, and can be explained by an active reaction driving the assembly.

In addition to the thermodynamic conditions, the cells also need to modulate the kinetics of condensation, to regulate when and where the assemblies would form. Many membraneless organelles such as nucleoli and histone-locus bodies form strictly at well-defined locations inside the cells, suggesting that the nucleation barrier is smaller at those select locations. Interestingly, the assembly of the phase separating components of nucleolus loses its spatiotemporal precision and {exhibits a high degree of spatiotemporal variability}  in the absence of such nucleolar organizer regions~\cite{Falahati2016277}. These nucleolar organizer regions are sites of ribosomal RNA transcription, and when activated, result in the transformation of the assembly process from a nucleation-limited homogeneous assembly to a growth-limited heterogeneous phase separation. {Therefore, in addition to the role of ribosomal RNA in modulating the thermodynamics of nucleolus assembly\cite{BerryPNAS2015}, it can also play a role in the spatiotemporal regulation of the nucleolus formation by changing the kinetics of assembly~\cite{Falahati2016277}.}

\section{Phase Separations in Biological Systems: Computational Investigations}
\label{section:computational}
\begin{figure*}
\begin{center}
	\includegraphics[width=.99\textwidth]{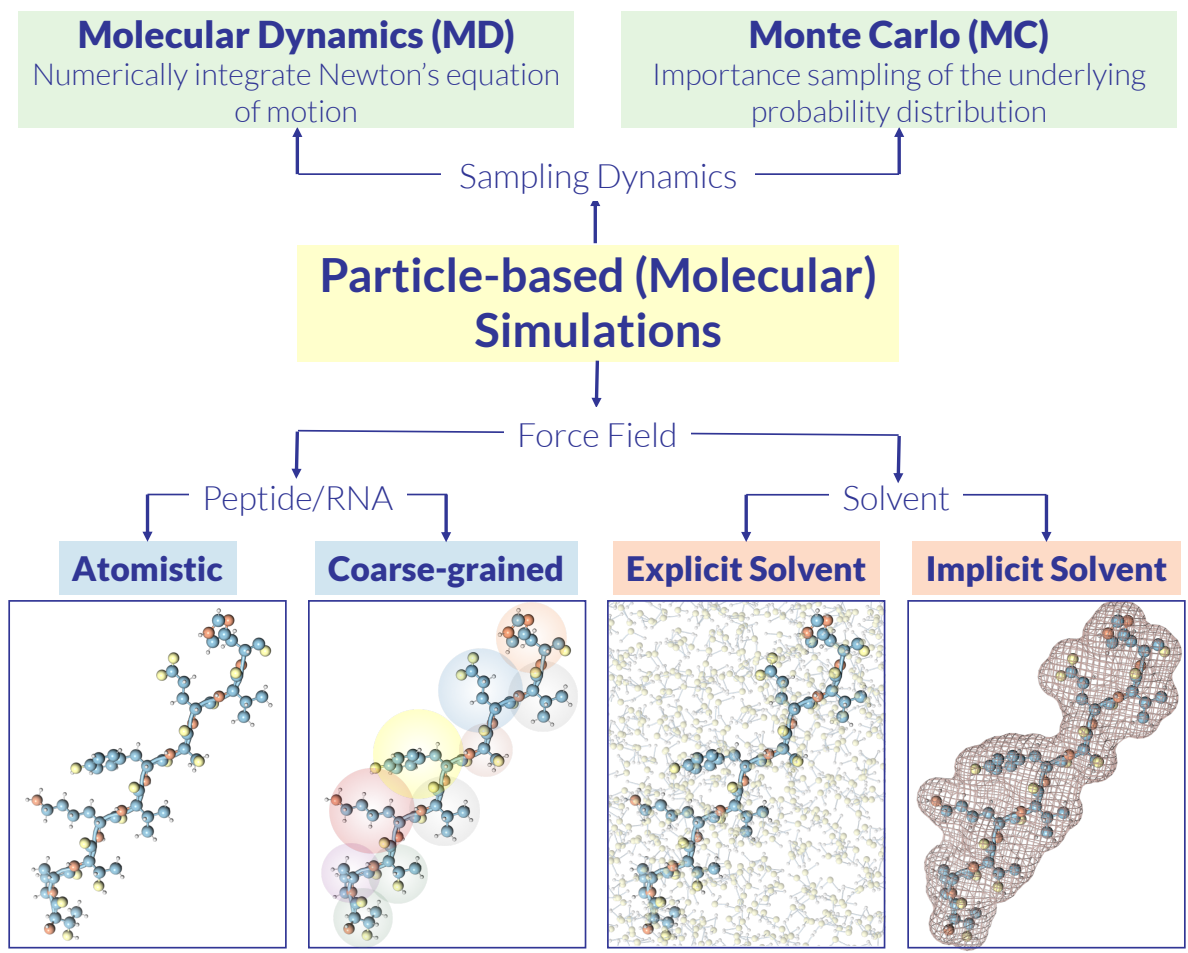}
	\caption{\label{fig:simulation-methodology}
	{Classification of particle-based (molecular) simulations of biological systems, based on the sampling dynamics and granularity of the utilized force-fields. Sampling the configuration space can be conducted using stochastic dynamics (in MC), or Newtonian dynamics (in MD). The utilized force-field for the biomolecule can be atomistic or coarse-grained. In both cases, solvent molecules can either be included explicitly in the simulation box, or their impact on the peptide/RNA chain can be represented using an effective field or modified interaction parameters. Coarse-grained models can have different levels of complexity, from each interaction site representing several atoms within a residue, to representing the entire macromolecule. }
	}
\end{center}
\end{figure*}

Experimental studies provide convincing evidence that thermodynamically driven LLPS processes play a significant role in  the assembly of membraneless organelles. The existing experimental techniques can indeed be utilized to characterize the thermodynamics and kinetics of LLPS both \emph{in vitro} and \emph{in vivo}. Their major disadvantage, however, is their inability to probe the molecular level events that culminate in phase separation, due to their limited spatiotemporal resolution. Similarly, experiments are not well equipped to probe the conformational ensembles and dynamics of IDPs and other phase separating proteins. In addition to these limitations, it is fairly expensive to use experiments for exploring large parameter spaces, e.g.,~assessing the sensitivity of phase separation propensity to changes in molecular properties and environmental conditions. Theoretical and computational approaches, however, do not have any of these limitations. They not only provide accurate molecular-level information about the conformation ensemble of proteins, and the molecular events that culminate in phase separation, but can also be used for large-scale sensitivity analysis studies. In addition, they can be utilized for probing length and timescales not otherwise accessible to experiments.

Theoretical and computational studies of LLPS can be broadly classified into two categories. In \emph{field-based} studies, which are usually continuum in nature, the effect of neighboring molecules on each molecule is represented by an effective field that depends on spatial profiles of thermodynamic variables, such as density and composition. The free energy of the system $\mathcal{F}[\textbf{c}(\textbf{r})]$ is then expressed as a functional of the density and/or composition profile $
\textbf{c}(\textbf{r})$, which is then minimized analytically or numerically using variational methods, subject to proper conservation constraints~\cite{CahnJChemPhys1958, CahnJChemPhys1959, CahnJChemPhys1959p}. Within field-based framework, demixing will occur under conditions at which a non-uniform composition profile minimizes the free energy functional. {Such functionals can also be utilized for constructing generalized partial differential equations for transport of individual components, making it possible to predict the spatiotemporal evolution of  composition in phase separating macroscopic and mesoscopic systems. }
Such descriptions have been extensively used to study flow~\cite{BadalassiJComputPhys2003} and phase separation~\cite{NaumanChemEngSci2001} in polymeric mixtures, and microphase separation in block copolymers~\cite{RenMacromolecules2001}. In general, these, alongside other flavors of field-based methods, such as the self-consistent field theory (SCFT)~\cite{FredricksonMacromolecules2002, FredricksonOxford2006}, have been an integral part of polymer physics in recent decades. Due to their simplicity and versatility, they are excellent tools for understanding the underlying physics of phase separation, and to assess how it is impacted by factors such as mixing enthalpy and surface tension. In recent years, field-based methods have been used for understanding LLPS in biological systems~\cite{ZwickerPNAS2014, BerryPNAS2015, MaoArxiv2018}.  When it comes to phase separation in a mixture of biomolecules with specific sequences, however, {standard} field-based methods have limited utility, as they lack the specificity and resolution needed to faithfully  account for specific interactions and spatial correlations present in such aqueous biomolecular solutions. 
{There have, however, been numerous attempts~\cite{LinPhysRevLett2016, LinJMolLiq2017, LinNewJPhys2017} in recent years to alleviate some of these limitations, via applying approaches such as the random phase approximation (RPA) method~\cite{BorueMacromolecules1988, GonzlezMozuelosJChemPhys1994}. Such approaches resolve heterogeneity at a single-chain level, but still  depend on the core assumption that single-chain characteristics  are predictive of phase separation propensity~\cite{LinBiophysJ2017}.   The applications of field-based approaches in understanding biomolecular phase separation has been the focus of several recent reviews~\cite{BerryRepProgPhys2018, LinBiochemistry2018, WeberArXiv2018}, and will not be discussed further in the current paper.} 

In \emph{particle-based} methods (Figure~\ref{fig:simulation-methodology}), which are the main focus of this section, a phase separating mixture is represented as a collection of particles, which interact according to a pre-determined potential energy function known as a \emph{force-field}. A force-field is usually  a linear combination of bonded (such as bonds, angles, dihedrals, etc) and non-bonded (e.g.,~Columbic, dispersion, etc) interactions. {
The primary advantage of particle-based methods is their ability to account for interatomic and intermolecular correlations without any \emph{a priori} assumptions about their nature, while in field-based methods, the mathematical form of such correlations-- or their lack thereof-- usually needs to be explicitly incorporated into the model.} Particle-based methods-- also known as \emph{molecular simulations}-- are classified into two categories. While \emph{molecular dynamics (MD)} simulations~\cite{AlderMDJCP1959} are based on integrating Newton's equations of motion, \emph{Monte Carlo (MC)} simulations~\cite{AlderWainwright1957} utilize an importance sampling stochastic scheme to generate a statistically representative sequence of configurations commensurate with the corresponding thermodynamic ensemble. Given sufficiently long sampling, both MC and MD are expected to yield identical averages for thermodynamic and structural properties if the underlying system is \emph{ergodic}. They might, however, be different in their efficiency, i.e.,~the statistical uncertainty of the desired thermodynamic averages vs.~the expended computational time. The major advantage of MD is its ability to predict dynamical properties, such as transport coefficients and nucleation kinetics, a task that cannot be achieved with MC due to lack of a rigorous mapping of MC moves to actual time. MC, however, is advantageous when the underlying force field is discontinuous, or when unphysical trial moves are utilized to enhance the sampling efficiency in slowly relaxing systems. Examples include particle swaps in multiphase systems or mixtures~\cite{PanagiotopoulosMolPhys1988}, and configurational bias approaches in simulations of polymers and peptides~\cite{SiepmannMolPhys1992}. Such unphysical moves are widely used in studying the conformational space of peptides, and the thermodynamics of biomolecular LLPS. 

Molecular simulations can also be classified based on the resolution of the utilized force-fields. The most detailed force-fields are \emph{atomistic} force-fields~\cite{JorgensenJACS1988, MacKerellBiopolymers2000, WangJComputChem2004, VegaPCCP2011}, in which every atom is represented by one-- or sometimes several-- interaction sites. {Atomistic simulations of biomolecular systems, however, are computationally expensive, and it is extremely difficult-- if not impossible-- to simulate systems with more than a million atoms, even with specialized hardware and massive parallelization\cite{PimptonLAMMPS1995, AndersonJCompPhys2008,  NguyenComputPhysComm2011}. There are two not mutually exclusive approaches for tackling these limitations. }\emph{Coarse-grained} force-fields expand the range of accessible length and timescales by averaging out atomistic details, and representing groups of atoms (e.g.,~residues in a peptide chain) as single interaction sites\cite{IzvekovJPhysChemB2005, ShellJChemPhys2008}. {In \emph{implicit-solvent} force-fields, which are used in both atomistic and coarse-grained simulations, solvent molecules are excluded, and their impact is accounted for with an effective field, or through modifying interaction parameters between the biomolecules~\cite{VitalisJCompChem2009}. As will be discussed later, coarse-grained force-fields can have different levels of granularity to the extent that a single interaction site can represent anything from several atoms within a residue, to the entire biomolecule.}

One of the major limitations of conventional molecular simulations-- irrespective of the utilized force-field-- is their inability to efficiently probe rare events, such as low-probability conformational rearrangements (e.g.,~protein folding) and nucleation. For that purpose, a wide range of \emph{advanced sampling techniques} have been developed, which expand the range of accessible timescales. Advanced sampling techniques can be classified into two categories. In \emph{bias-based} methods, such as umbrella sampling~\cite{TorrieJCompPhys1977}, flat histogram methods~\cite{WangPhysRevLett2001, ShellPhysRevE2002} and metadynamics~\cite{LaioPNAS2002, BarducciPRL2008} the underlying Hamiltonian is modified in order to preferentially favor certain configurations. In \emph{path sampling methods}, such as transition path sampling~\cite{ChandlerTPS2002}, forward-flux sampling~\cite{AllenPhysRevLett2005, FrenkelFFS_JCP2006, HajiAkbariJChemPhys2018}, transition interface sampling~\cite{vanErpJChemPhys2003, BolhuisJCompPhys2005} and parallel tempering~\cite{SwendsenPRL1986, EarlPCCP2005, RathoreJChemPhys2005}, the underlying Hamiltonian is not modified, and trajectories are instead sampled in a targeted manner. 

The remainder of this section will be dedicated to discussing the types of information that molecular simulations can provide about LLPS in biological systems. The studies discussed here utilize a wide variety of simulation techniques, and can be conceptually categorized into two groups. The first group deals with IDPs as main culprits for LLPS in biological systems, and explores their thermodynamics, conformational ensembles and dynamics. The second group, however, directly deals with the question of biological liquid-liquid phase separation, and understanding its thermodynamics and kinetics. 

\subsection{Computational Investigation of IDPs, the Main Culprits of Biological LLPS}\label{section:IDP-simulations}
\label{section:computational:IDPs}
While IDPs and proteins with IDRs have been known for a long time, the origin of their {disorder still remains elusive. This question is closely related to a decades-old problem in polymer physics, i.e.,~the question of what dictates a polymer's conformation in solution.} The conditions under which a polymer will collapse onto a compact globule have been extensively studied, and it is generally understood that it is rare for a heteropolymer with a random sequence of solvophobic and solvophillic residues to simultaneously fold into a globular structure and remain solvated~\cite{KauzmannAdvProteinChem1959, FisherPNAS1964}. In order for such folding to occur, a tight balance needs to be established between the entropic loss due to adopting a folded structure, and the enthalpic gain due to favorable energetic interactions among the core solvophobic residues, and among the solvophilic residues and the solvent. Such a condition is usually not met for a random heteropolymer~\cite{DillBiochemistry1985}. Peptides with well-defined folded structures therefore constitute remarkable anomalies. This is not surprising as naturally occurring peptides are not random heteropolymers, and instead have sequences naturally selected through evolution. Understanding the relationship between sequence and folding propensity has been the focus of multiple studies. In a pioneering work,  Uversky~\emph{et al.}~\cite{UverskyProteinsStructFunctGenet2000} demonstrated  that the propensity of a peptide to form well-defined folded structures is dictated by its mean per-residue net charge, and the average hydrophobicity of its constituent residues, and the fact that IDPs do not have well-defined folded structures is due to their high net charge and low hydrophobicity. This minimal model, however, cannot be fully predictive, as folding propensity can be strongly impacted by subtle features such as the linear distribution of charged residues along a sequence~\cite{vanDerLeeChemRev2014}. Therefore, even though sequence-based methods for predicting folding propensity have become increasingly accurate over years~\cite{RomeroApplBioinfo2004, DosztanyiBioinformatics2005, DosztanyiBriefBioinform2009, MonastyrskyyProteinsStructFunctGenet2011, MonastyrskyyProteinsStructFunctGenet2014}, they are not well-equipped to properly account for all such nuances. Moreover, such approaches are usually incapable of providing molecular-level information about the conformational dynamics of the peptides that they might correctly detect as intrinsically disordered. 

In recent decades, molecular simulations have emerged as attractive tools for inspecting the origin of IDP disorder, their conformational dynamics and their binding propensity.  The predictive power of such studies were, however, limited until recently, primarily due to the tendency of most classical protein force fields to overestimate the formation of secondary structures.  This problem is indeed universal across the board, and affects both explicit- and implicit-solvent atomistic and coarse-grained force-fields. This issue has, however, been addressed~\cite{BestCurrOpinStrucBiol2017} in more recent all-atom force-fields, e.g.,~by strengthening water-protein interactions~\cite{NerenbergJPhysChemB2012, BestJChemTheoryComput2014, HuangNatMethods2016} and utilizing more realistic water models~\cite{PianaJPhysChemB2015}. Also, newer force-fields, such as the Kirkwood-Buff force field (KBFF)~\cite{MercadanteJPhysChemB2015}, have been developed with a focus on solution properties. Similar improvements have been made to implicit-solvent~\cite{VitalisJCompChem2009} and coarse-grained~\cite{DignonPLOSComputBiol2018} models. 

\begin{figure}
	\centering
	\includegraphics[width=.4\textwidth]{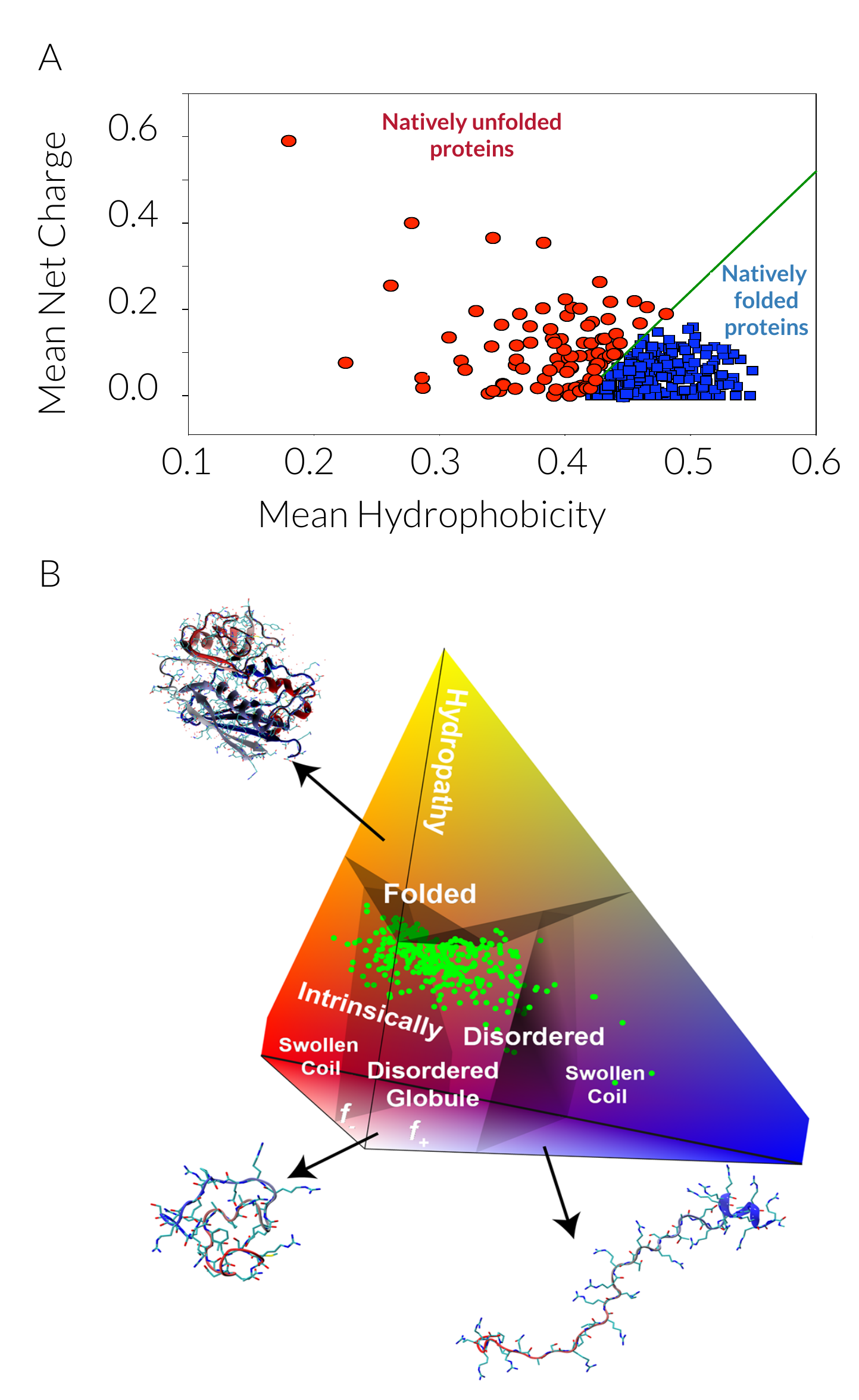}
	\caption{{(A) (Adapted and relabeled from Ref.~\citenum{DunkerBMCGenomics2008}) The original hydropathy/mean charge correlation of Uversky~\emph{et al}. (B) (Reproduced from Ref.~\citenum{vanDerLeeChemRev2014}) The three-dimensional correlation of Mao~\emph{et al.} that describes the behavior of IDPs based on the fraction of positively and negatively charged residues. }}
\end{figure}

{One of the most pressing question about IDPs is the extent to which their disorderedness is impacted} by factors such as amino acid sequence, temperature, and the presence of inert crowders. A powerful way to address these questions is to utilize highly coarse-grained force-fields in which distinct amino acids with similar properties (e.g.,~charge, hydrophobicity) are encoded into a single type of interaction site. {Despite not being} sequence-specific, {such models} qualitatively capture the underlying physics of interactions between a peptide and solvent molecules. One major advantage of using such models is the computational efficiency at which ''generic`` peptides with a sufficiently wide range of desired solvation and electrostatic properties can be simulated.  For instance, the original idea of Uversky~\emph{et al.}~\cite{UverskyProteinsStructFunctGenet2000} was explored by Ashbaugh and Hatch~\cite{AshbaughJACS2008} using a simple bead-spring model of peptides in which each residue can only be of three types: hydrophobic, polar, and positively charged polar. They simulated sequences with a wide range of lengths ($N$), net charges $(q)$ and hydrophobicities ($H$), with 15-20 random peptides generated for each $N$, $q$ and $H$. Even though they observed that coil-like and collapsed globules occupy distinct regions of the $(q,H)$ space, the boundary between them only becomes $N$-independent (as predicted by Uversky) when explicit counterions are included for charged residues (vs.~utilizing screened charges). A similar approach was utilized by Miller~\emph{et al.}~\cite{MillerBiophysJ2016} for probing the impact of crowders on the compactness of peptides (e.g.,~their radii of gyration). In the presence of inert (i.e.,~repulsive) crowders, they observed that coil-like peptides become more compact, while no significant increase in compactness was observed for globular peptides. In other words, crowding can be used as a means of modulating IDP compactness in biological systems. {Note that it is not feasible to use experiments or atomistic molecular simulations to explore the effect of crowders on the conformational ensembles of IDPs, and this is a question that can only be addressed using such highly coarse-grained models.}

Despite their usefulness in elucidating the qualitative behavior of IDPs, such coarse-grained models are not suitable for making quantitative predictions about the conformational dynamics and binding propensities of {a particular IDP}. Such detailed information about IDPs can usually be obtained from atomistic simulations only, which, thanks to recent advancements in computer architecture, have become faster and more popular in recent years~\cite{DasACSCentSci2018}.  For instance, numerous atomistic studies have been conducted with the aim of understanding the effect of charge distribution on disorder propensity. As an example, Mao~\emph{et al.} utilized the ABSINTH implicit-solvent model~\cite{VitalisJCompChem2009} and MC simulations to explore the conformational ensemble of peptides with low hydrophobicity~\cite{MaoPNAS2010}. As predicted  by the  Uversky~\emph{et al.}'s hydrophobicity/mean net charge correlation~\cite{UverskyProteinsStructFunctGenet2000}, such peptides tended to remain unfolded. Depending on the fraction of positively and negatively charged residues, however, they exhibited swollen coil, or disordered globule conformations. In other words, different IDPs can have different levels of chain compactness, which is key to their biological function. IDP conformation can also be impacted by charge decoration, or the linear distribution of positively and negatively charged residues along the sequence, as demonstrated in a recent work by Firman and Ghosh\cite{FirmanJChemPhys2018}. They derived an analytical theory for coil-to-globule transition in heteropolymers, and calibrated it against findings of atomistic MC  simulations of IDPs in the DisProt~\cite{SickmeierNucleiAcidsRes2006} database. They observed that IDPs with identical $f_+$'s and $f_-$'s can exhibit different conformations, depending on factors such as charge decoration, and post-translational modifications. {The role of charge decoration on phase separation propensity has also been demonstrated in on-lattice MC~\cite{DasJPhysChemB2018} and coarse-grained MD~\cite{DasPCCP2018} simulations.} Note that it is nontrivial to account for such effects in sequence-based methods, demonstrating the importance of atomistic simulations in accessing the conformational ensemble of IDPs.

An interesting phenomenon that has been recently studied using molecular simulations is the collapse of some unfolded peptides-- including some IDPs-- upon increasing temperature~\cite{SadqiPNAS2003, NettelsPNAS2009, LangridgeProteinsStructFunctGenet2014}. This behavior is in contrast to the tendency of most polymers to swell upon heating~\cite{SanchezMacromolecules1979}, and has been attributed to temperature-dependent per-residue hydration energies for such peptides~\cite{WuttkePNAS2014, ZerzeJPhysChemB2015, HicksJChemPhys2018}. For instance, Wuttke~\emph{et al.}~\cite{WuttkePNAS2014} investigate chain compaction in five proteins that can exist in an unfolded state under physiologically relevant conditions using FRET experiments, theory and molecular simulations. Using a theoretical description proposed by Sanchez~\cite{SanchezMacromolecules1979}, they explained the observed heat-induced compaction using a $T$-dependent effective monomer-monomer interaction parameter. They also conducted molecular simulations using the implicit-solvent ABSINTH model, and confirmed that the heat-induced chain compaction will only occur if the temperature dependence of per-residue hydration energies are explicitly included in the  ABSINTH model. Later studies using explicit-solvent force-fields confirm such compaction~\cite{ZerzeJPhysChemB2015, HicksJChemPhys2018}, and the-dependence of solvent accessible surface area (SASA) on temperature~\cite{ZerzeJPhysChemB2015}. From a molecular perspective, this effect can be attributed to the preponderance of nascent structures in unfolded peptides, which makes them distinct from random-coil polymers~\cite{ZerzeJPhysChemB2015, HicksJChemPhys2018}. Note that it is extremely difficult to attain such granular residue-level information using experiments, and yet such information are pivotal for understanding IDP properties and function~\cite{ZerzePhysRevLett2016, LevineCurrOpinStructBiol2017}. Due to the rugged free energy landscape of the large peptides studied in Refs.~\citenum{WuttkePNAS2014, ZerzeJPhysChemB2015, HicksJChemPhys2018}, advanced sampling techniques such as replica exchange molecular dynamics~\cite{MittalJPhysChemB2012, ZerzeJCTC2015, ZerzeJPhysChemB2015p} were utilized for generating statistically representative peptide conformations.

Molecular simulations have also proven useful in exploring the effect of probes and chromophores on the conformational ensembles and dynamics of IDPs, an information critical to interpreting data obtained from probe-based techniques. One of these techniques is the F\"{o}rster resonance energy transfer (FRET)~\cite{ForsterAnnPhys1948} in which two chromophores, one energy donor and one energy acceptor, are covalently placed at select locations along a peptide (usually at the C and N termini). Energy is then transferred from the excited donor to the ground-state acceptor through dipole-dipole coupling. The efficiency of such energy transfer will be proportional to $r^{-6}$, with $r$ the distance between the two probes. FRET energy transfer efficiency is therefore a sensitive measure of the distance between the two probes, and is widely used for exploring the conformational ensemble and dynamics of IDPs. In order to properly interpret FRET findings, it is necessary to make sensible assumptions about the impact of probes on the conformational ensemble, dynamics, and long-range contact distribution of IDPs. In recent years, this question has been extensively studied using molecular simulations~\cite{ZerzeBiophysJ2014, SmithPhysRevE2014, ZhengJACS2016, YooELife2018, LiMolBiosys2016,  ZhengJChemPhys2018}.  For instance, Zerze~\emph{et al.} investigated\cite{ZerzeBiophysJ2014} the effect of fluorophores on the compactness, secondary structure content, and long-range contact distribution of three unfolded proteins (the cold shock protein, CSP) from \emph{Thermotoga maritima}, the DNA-binding domain of l-repressor (LR), and the N-terminal domain of HIV integrase (IN), using replica exchange MD\cite{SugitaChemPhysLett1999}, and they did not find such structural metrics to be strongly impacted by the fluorophores utilized in FRET experiments. Since then, numerous other computational studies have attempted to characterize different aspects of FRET experiments~\cite{SmithPhysRevE2014, ZhengJACS2016, YooELife2018, LiMolBiosys2016,  ZhengJChemPhys2018}.

\subsection{Computational Investigation of Biological LLPS}
\label{section:computational:LLPS}

The computational studies discussed in Section~\ref{section:IDP-simulations} involve simulating one (or a handful of) IDP chains, and can therefore be conducted using fully atomistic force-fields.  It is, however, far more expensive computationally to study the thermodynamics and kinetics of LLPS for the following reasons. First of all, it is not generally possible to obtain a realistic picture of LLPS without simulating systems comprised of at least several hundred IDPs (or other phase separating biomolecules), a daunting and computationally intractable task if atomistic force-fields are to be utilized. Developing accurate coarse-grained force-fields of proteins is therefore a prerequisite to studying the thermodynamics and kinetics of LLPS in biological systems. Secondly, depending on the magnitude of the thermodynamic driving force, LLPS can be nucleation-limited. Crossing the nucleation barrier will be a rare event under such circumstances, and will occur over timescales much longer than what would take for the conformational rearrangement of individual chains. As discussed earlier, a wide variety of advanced sampling techniques have been developed for studying rare events in recent decades, which need to be used for studying the kinetics and free energy landscape of nucleation-limited LLPS in biological systems.

Due to these complexities, very few computational studies of the actual LLPS process have been conducted. Most such studies employ highly coarse-grained implicit-solvent nonspecific models of proteins, and are thus only designed to provide qualitative information about the thermodynamics and kinetics of phase separation in protein-like systems.  Historically, studies of biological phase separation predate recent interest in LLPS, and, like experimental efforts, were originally conducted with the aim of understanding the crystallization of globular proteins. In such representations, individual proteins are treated as colloidal particles interacting via isotropic~\cite{tenWoldeScience1997} or patchy~\cite{DoyePCCP2007, GogeleinJChemPhys2008} potentials. The interest in probing the thermodynamics and kinetics of phase transitions in patchy colloidal systems~\cite{ZhangNanoLett2004, BianchiPhysRevLett2006, RomanoNatMater2011, SmallenburgNatPhys2013, JacobsJChemPhys2014, SimonBioArxiv2018}, however, extends far beyond understanding protein crystallization, as patchy colloids are excellent model systems for inspecting how a competition between isotropic and directional interactions can impact the phase behavior, e.g.,~the relative stability of different crystals and liquids, as well as the kinetics of self-assembly.

In recent years, several qualitative studies of biological LLPS have been conducted, mostly employing simple on-lattice models. For instance, Jacobs and Frenkel~\cite{JacobsBiophysJ2017} used grand canonical Monte Carlo (GCMC)~\cite{AdamsMolPhys1974} to simulate an $N$-component lattice gas, in which the contribution of two adjacent occupied sites of types $i$ and $j$ to the total energy was given by $\epsilon_{ij}$.  Inspired by the pioneering theoretical work of Sear and Cuesta~\cite{PhysRevLett.91.245701}, who had used random matrix theory to identify conditions for LLPS in many-component biological systems, they drew  $\epsilon_{ij}$'s from a Gaussian distribution of pre-specified  mean and variance, and observed that de-mixing will occur if the standard deviation in interaction strength exceeds a well-defined threshold. This study provides some clarity to the otherwise hopeless question of LLPS \emph{in vivo}, and stipulates that the propensity of a complex biomolecular mixture to undergo LLPS can be tuned by changing the variance of interaction strength between its constituent components. 

		Despite their utility in providing insight into how demixing propensity is impacted by factors such as interaction strength, such coarse-grained models cannot be quantitatively predictive when a particular protein is concerned, and are particularly unsuitable for studying IDPs due to their poorly defined generic geometries and conformational ensembles.  In recent years, however, there have been major advancements~\cite{GhavamiJCTC2012, DignonPLOSComputBiol2018} in developing accurate and computationally efficient coarse-grained protein force-fields in which each amino acid is represented using a single interaction site, and solvent molecules are only considered implicitly. Also, charged residues that are prevalent in IDPs interact via screened electrostatic potentials~\cite{RudisillMolPhys1989, SmitMolPhys1991}, as predicted by the Debye-H\"{u}ckel theory~\cite{DebyePhysZ1923}. Therefore, due to the short-range nature of such screened electrostatic interactions, there is no need to utilize computationally costly methods such as the Ewald sum~\cite{DeLeeuwProcRSocLondon1980}. This latter fact makes such coarse-grained representations even faster. It is therefore now  feasible to simulate hundreds of IDP chains on existing computer architectures. This has resulted in a few studies of direct coexistence of phase separating IDPs. For instance, Dignon~\emph{et al.}~\cite{DignonPLOSComputBiol2018} utilize the slab method~\cite{BlasJChemPhys2008, SilmoreMolPhys2017} to study the phase separation of two proteins, i.e.,~the low complexity domain of the RNA-binding protein FUS and the DEAD-box helicase protein LAF-1. In the slab method, the coexisting densities and compositions are accurately determined by placing a slab of the peptide-rich phase in contact with vacuum, which represents the solvent-rich phase. By doing so, the thermodynamics of phase separation can be systematically investigated without a need to cross the nucleation barrier. The question of interest in Dignon~\emph{et al.}'s work, in particular, was to understand how LLPS is qualitatively affected by FUS mutations, and the presence-- or lack thereof-- folded domains in LAF-1. In a separate study\cite{DignonPNAS2018}, they utilized their coarse-grained force-field alongside the slab method to demonstrate that there is a close correlation between $T_c$, the critical demixing temperature, $T_\theta$ the temperature at which coil-to-globular transition occurs for an isolated chain, and $T_B$, the Boyle temperature at which the second virial coefficient vanishes. They established such correlation by simulating 20 different IDP sequences with a wide range of properties. This is a very important finding and demonstrates that properties such as $T_B$ and $T_\theta$ that can be determined from observing the behavior of isolated chains can be predictive of the propensity of a dense mixture of such chains to undergo phase separation. {These findings prove the applicability of the random phase approximation method that assumes the existence of such a correlation\cite{LinBiophysJ2017}.}

\section{Emerging Questions and Path Forward}
\label{section:outlook}
Thermodynamically driven phase separations provide an energetically inexpensive means for cells to concentrate certain proteins and RNAs. Such structures can be liquid-like or solid-like, which determines the types of  biological functions that they undertake.
For instance, the fact that membraneless organelles are liquid-like allows for fast diffusion of components including substrates and products in and out of such assemblies. Crystalline or amorphous solid assemblies, however, can provide structural support to the cell. They can also attract and retain certain molecules from the pool of readily available cellular components, e.g.,~in the case of stress. Over all, the phase separation model has been successful in explaining certain aspects of the assembly and function of membraneless organelles. Yet, the complexity of biological systems makes it extremely difficult to validate and characterize the occurrence of LLPS processes \emph{in vivo}.

One such complexity arises from the active nature of living organisms. Despite  growing evidence in recent years in favor of the LLPS model, the contribution of active reactions to the formation and disassembly of membraneless organelles \emph{in vivo} cannot be disregarded\cite{doi:10.4161/nucl.2.3.16246, 10.7554/eLife.04591, Kedersha1431, Jain:aa,BUCHAN20131461, loschi2009dynein, Thomas2011324, doi:10.1021/acs.biochem.8b00025}. For instance, depletion of ATP completely blocks the assembly of stress granules, whose proteome contains a large number of ATP-dependent helicases and protein remodelers~\cite{Jain:aa}. {In principle, the large number of active reactions, such as RNA transcription, that occur within membraneless organelles, can be sufficient to locally concentrate macromolecules, and drive the assembly of membraneless organelles\cite{Dundr01122010, Shevtsov:2011aa, karpen1988drosophila, Grob01022014, Salzler:aa}. For instance, the local accumulation of the products of such functional reactions can drive the recruitment of their binding partners, such as processing factors, a process that does not need to be a thermodynamically driven phase separation\cite{BerryPNAS2015, Falahati2016277}. Another active process that can culminate in the assembly of membraneless organelles is the active transport of their constituent components.  For instance, the formation of stress granules and the growth of P-bodies in response to stress, rely on motor proteins \cite{loschi2009dynein,Thomas2011324}. Interestingly, the formation of high concentration assemblies with liquid-like properties have also been observed in active swimmers whose motion is powered by a chemical reaction even in the absence of any attractive forces\cite{TAKATORI201624}. Finally,  biochemical processes such as transcription, translation, and post-translational modifications, that result in the formation of the components of membraneless organelles, are all comprised of active chemical reactions that will proceed only in the presence of ATP.}
While the contribution of such active processes to the formation of membraneless organelles can be through modulation of LLPS, they can also function as the sole driving force for the formation of membraneless organelles.  In other words, while each model can explain certain aspects of the membraneless organelle assembly, distinguishing between a solely-active assembly mechanism, and a combination of active and LLPS is particularly challenging and requires a more extensive thermodynamic characterization of the macromolecules that localize to such organelles \emph{in vivo}.
{In recent years, numerous phenomenological continuum models  have been proposed for understanding the role of reactions and concentration gradients in phase separation\cite{HymanRev, LeePhysRevLett2013, ZwickerNatPhys2017, WeberNewJPhys2017, WurtzNewJPhys2018,  ZwickerPhysRevLett2018}.}  Developing additional testable theoretical models based on the predictions of the two mechanisms are pivotal for discerning the potential contribution of each mechanism to a particular assembly process.

A second complication arrises from the complexity of biomolecules, and the difficulty of probing all timescales relevant to their conformational rearrangements and assembly. {Experimentally, this can, in principle, be addressed by using probe-based techniques such as FRET\cite{Elbaum-Garfinkle09062015, NottNatChem2016, Monahane201696394}, or nuclear magnetic resonance (NMR) spectroscopy\cite{Burke2015231, Conicella_2016, Monahane201696394}. Unfortunately, these techniques do not have the sufficient resolution for realtime probing of biomolecular conformations. Developing advanced spectroscopy and microscopy techniques with increased resolution will be particularly helpful in this regard. One possible avenue is to re-engineer techniques such as femtosecond ultrafast scattering\cite{GaffneyScience2007, ChapmanNature2011} and four-dimensional electron microscopy\cite{ZewailScience2010} to be compatible with biomolecular systems.  In addition, }this is an area that can benefit immensely from molecular simulations. Despite major breakthroughs in recent decades, we are only witnessing the beginning of using molecular simulations to investigate IDPs and biological LLPS. With more accurate force-fields, more accurate and efficient advanced sampling techniques, and better and faster hardware, it is now possible to investigate conformational rearrangements of a wide range of proteins, as well as their propensity to bind other proteins and ligands. Moreover, recent advancements in systematic coarse-graining have made it feasible to explore biological LLPS in real time. In particular, using such coarse-grained force-fields in conjunction with advanced sampling techniques can provide us with accurate information about the kinetics and mechanism of phase separation, and thus help us address deep fundamental questions about proteins and protein folding and assembly. One of the most pressing questions is the role and importance of hydrophobicity in inducing biological self-assembly. This question has fascinated statistical physicists and biophysicists for decades~\cite{KauzmannAdvProteinChem1959, DillBiochemistry1990, ChandlerNature2005}, and has been recently investigated for simpler hydrophobic models using extensive molecular simulations and path sampling techniques~\cite{PatelPNAS2011, SumitPNAS2012, AltabetJCP2014, RemsingPNAS2015,  AltabetPNAS2017}. Nevertheless, the precise kinetics and mechanism of hydrophobic assembly in biological  systems is far from fully understood, and considering the power of state-of-the-art advanced sampling techniques in elucidating the mechanism of other rare events such as crystallization~\cite{LiNatMater2009, GalliNatComm2013, HajiAkbariFilmMolinero2014, HajiAkbariPNAS2015, CabrioluPRE2015, GianettiPCCP2016,  HajiAkbariPNAS2017}, they can be very useful in understanding the hydrophobic effect. IDPs are excellent systems in this regard, as they tend to phase separate and assemble despite having low hydrophobicity, and understanding the molecular origins of such behavior is critical to unraveling the longstanding question of hydrophobicity and biological assembly.

{Finally, an important challenge in studying LLPS in biological systems is the difficulty of controlling thermodynamic variables (such as temperature, concentration, pH) and operating conditions (such as hydrodynamic shear) at a cellular level. Such a challenge not only makes it difficult to obtain reliable data about the kinetics of biomolecular assembly \emph{in vivo}, but also makes a robust analysis of the sensitivity of LLPS to such factors challenging. In recent years, optogenetic and microfluidic approaches have been successfully utilized to address this issue. For instance, optogenetic approaches can be used for spatial modulation of concentration locally and globally within cells\cite{ShinCell2017, BrachaCell2018}. Another powerful approach that has gained increased popularity in the soft matter community in recent years is microfluidics, which allows for swift and precise control of operating variables such as temperature, pressure, concentration, pH and shear deformation. This feature makes microfluidics particularly ideal for rate measurements, as they have been successfully used for probing nucleation kinetics in several systems\cite{ChenJACS2005, GerdtsAngewChemIntEd2006, ShimCrystGrowthDes2007, WhitesidesNature2006}. Microfluidic devices were recently employed for studying the impact of rapid changes in temperature on the \emph{in vivo} assembly of nucleolar proteins in \emph{Drosophilla} embryos\cite{Falahati07022017}. Developing similar experimental techniques that will enable us to explore the molecular driving forces and kinetics of LLPS can be another potential area of future exploration.}

\acknowledgements
A.H.-A. gratefully acknowledges the support of the National Science Foundation CAREER Award (Grant No. CBET-1751971). H.F. is a Howard Hughes Medical Institute Fellow of the Life Sciences Research Foundation (LSRF).

\bibliographystyle{apsrev} 
\bibliography{ReviewRefs.bib} 

\end{document}